\documentstyle[aps,prl,epsf]{revtex}

\def\lsim{\raise0.3ex\hbox{$<$\kern-0.75em\raise-1.1ex\hbox{$\sim$}}}
\def\gsim{\raise0.3ex\hbox{$>$\kern-0.75em\raise-1.1ex\hbox{$\sim$}}}

\begin{document}

\draft

\twocolumn[\hsize\textwidth\columnwidth\hsize\csname
@twocolumnfalse\endcsname
\title{
\vspace*{-15pt}
{\normalsize \hfill {\sf UTCCP-P-85}} \\
Dynamical Quark Effects on Light Quark Masses}

\author{A.~Ali Khan$^1$,
        S.~Aoki$^2$,
        G.~Boyd$^1$,
        R.~Burkhalter$^{1,2}$,
        S.~Ejiri$^1$,
        M.~Fukugita$^3$,
        S.~Hashimoto$^4$
        N.~Ishizuka$^{1,2}$,
        Y.~Iwasaki$^{1,2}$,
        K.~Kanaya$^{1,2}$,
        T.~Kaneko$^4$,
        Y.~Kuramashi$^4$,
        T.~Manke$^1$,
        K.~Nagai$^1$,
        M.~Okawa$^4$,
        H.P.~Shanahan$^1$,
        A.~Ukawa$^{1,2}$ and
        T.~Yoshi\'e$^{1,2}$\\
	(CP-PACS Collaboration)}
\address{
$^1$Center for Computational Physics,
University of Tsukuba, Tsukuba, Ibaraki 305-8577, Japan \\
$^2$Institute of Physics, University of
Tsukuba, Tsukuba, Ibaraki 305-8571, Japan \\
$^3$Institute for Cosmic Ray Research,
University of Tokyo, Tanashi, Tokyo 188-8502, Japan \\
$^4$High Energy Accelerator Research Organization
(KEK), Tsukuba, Ibaraki 305-0801, Japan}

\date{April 2000, revised July 2000}
\maketitle

\begin{abstract}\noindent
We present results for light quark masses from a systematic lattice QCD 
study with two degenerate flavors of dynamical
quarks. Simulations are made with a renormalization-group 
improved gauge action and a
mean-field improved clover quark action for sea quark masses corresponding to
$m_{\rm PS}/m_{\rm V} \approx 0.8$--0.6 and the lattice spacing
$a=0.22$--0.11 fm.  In the continuum limit we find 
$m_{ud}^{\overline{MS}}(2\,{\rm GeV})=3.44^{+0.14}_{-0.22}$ MeV
using the $\pi$ and $\rho$ meson masses as physical input, 
and $m_s^{\overline{MS}}(2\,{\rm GeV})=88^{+4}_{-6}$ MeV or $90^{+5}_{-11}$
MeV with the $K$ or $\phi$ meson mass as additional input.
The quoted errors represent statistical and systematic combined, 
the latter including those from continuum and chiral extrapolations, 
and from renormalization factors. Compared to quenched results, two flavors of 
dynamical quarks reduce quark masses by about 25\%. 
\end{abstract}

]

\narrowtext

Masses of light quarks belong to the most fundamental parameters of the
Standard Model\cite{weinberg}, and yet their precise values have been 
difficult to determine due to quark confinement.  
Lattice QCD provides a fundamental approach
to overcome this problem\cite{reviews}
since it enables first principle calculations of hadron masses as functions 
of quark masses.  
This approach has progressed considerably recently\cite{fnal,cp-pacs,recent} 
through high statistics calculations that have allowed
the continuum limit to be taken for quark masses, 
and the development of non-perturbative renormalization 
techniques for a reliable conversion of lattice bare quark masses to 
those in the continuum.

These studies, however, have been 
carried out within the quenched approximation which ignores 
effects of sea quarks.
A limitation, shown in Ref.\cite{cp-pacs}, 
is that the strange quark mass 
cannot be consistently determined,  
with the values differing by 20\% depending on 
the choice of meson mass taken for input. 
It has been suspected\cite{reviews,fnal}, furthermore,
that dynamical sea quark effects sizably 
reduce the values of light quark masses.

Clearly, systematic full QCD studies incorporating dynamical 
sea quark effects are needed for progress in the determination of quark 
masses.  A recent attempt has been reported in\cite{SESAMquark}. 
In this Letter we present results of our investigation
in which the $u$ and $d$ quarks, assumed degenerate, are simulated 
dynamically while the $s$ quark is treated in the quenched 
approximation\cite{preliminary}. 
 
Full QCD simulations are computationally much more demanding
than quenched simulations.
This problem can be significantly eased by the use of improved actions.
Because of reduced cutoff errors, 
they should allow continuum extrapolations 
from coarser lattices, and hence 
require smaller lattice sizes, and smaller computational costs, 
for simulations with the same physical lattice size. 

We employ improved actions both for the gluon part and the quark part. The
gluon action consists of $1\times 1$ and $1\times 2$ Wilson loops whose
coefficients are determined by an approximate renormalization-group
analysis\cite{iwasaki}.  For the quark part we choose the ``clover''
improvement of the Wilson action\cite{clover}, adopting, for the clover
coefficient, a mean-field value $c_{SW} = P^{-3/4}$. We substitute the
one-loop result $P=1-0.1402g^2$\cite{iwasaki} for the plaquette $P$, which
agrees within 8\% with the values measured in our runs. The one-loop result
of $c_{SW}$\cite{aoki} is found to be close to our choice.

The improved action described here was tested in our
preparatory full QCD study\cite{compara}.  
We found that scaling violation in hadron masses is small 
with this action already at $a^{-1}\approx 1$~GeV, 
as compared to $a^{-1} \gsim\, 2$~GeV needed  
for the standard plaquette and Wilson quark actions.   
We therefore aim at a continuum extrapolation from simulations 
made at $a^{-1}\approx 1$--2~GeV.

We make runs at three values of the coupling 
$\beta\equiv 6/g^2 = 1.8, 1.95, 2.1$ to cover this range, 
employing lattices of a similar physical spatial size $La \approx 2.5$~fm, 
as listed in Table~\ref{tab:param}.
For each $\beta$, gauge configurations are generated by the 
hybrid Monte Carlo algorithm at four values of the sea quark hopping 
parameter $\kappa_{sea}$ corresponding to the pseudoscalar (PS) to vector (V)
meson mass ratio of $M_{PS}/M_V \approx 0.8$, 0.75, 0.7 and 0.6. 
For each sea quark mass, 
we calculate hadron masses at five values of the valence quark 
hopping parameter $\kappa_{val}$ corresponding to 
$M_{PS}/M_V \approx 0.8$, 0.75, 0.7, 0.6 and 0.5, taking unequal
as well as equal quark mass cases. 
Masses are extracted from hadron propagators
with the standard correlated $\chi^2$ fit. Errors are estimated with the
jackknife procedure with a bin size of 50 trajectories, derived
from an autocorrelation study.

For the Wilson-type quark action including the clover action, different
definitions of quark masses lead to results that differ at finite lattice
spacing due to explicit breaking of chiral symmetry.
We employ three definitions in the present work, checking consistency 
among them for reliability of results:    
(i) using the axial vector Ward identity, we define,
\begin{equation}
m_q^{\rm AWI}a 
=  \lim_{t\to\infty}\frac{\langle\nabla_4 A^{\rm imp}_4(t)P(0)\rangle}
			{2\langle P(t)P(0)\rangle}, 
\label{eq:awi-mass}
\end{equation}
with $P$ the pseudoscalar density and $A^{\rm imp}_4=A_4+c_A\nabla_{4}P$ 
the axial vector current improved to $O(a)$; 
(ii) another possibility, suggested by the vector Ward identity 
and naturally appearing in perturbative analyses, reads 
$m_q^{\rm VWI}a =(1/\kappa-1/\kappa_c)/2$ where $\kappa_c$ represents 
the critical hopping parameter at which the PS meson mass vanishes
$M_{PS}(\kappa_{val}=\kappa_{sea}=\kappa_c)=0$;  
(iii) a third possibility, suggested in\cite{gupta-bhatta} and denoted by
$m_q^{\rm VWI,PQ}$, replaces 
$\kappa_c$ in (ii) by the ``partially quenched'' critical value 
$\kappa_c^{\rm PQ}$ where the PS meson mass vanishes 
as a function of the valence hopping parameter $\kappa_{val}$  
when $\kappa_{sea}$ for sea quark is fixed to the physical point 
of $u$ and $d$ quark,
$M_{PS}(\kappa_{val}=\kappa_c^{\rm PQ};\kappa_{sea}=\kappa_{ud})=0$.

We express the PS meson mass $M_{PS}^2$ in terms of quark masses 
by a general quadratic ansatz of the form, 
\begin{eqnarray}
M_{PS}^2a^2 &=&  b_s m_{sea}a + b_v m_{val}a 
	        +c_s (m_{sea}a)^2 \nonumber\\
            &+&  c_v (m_{val}a)^2 
                +c_{sv} m_{sea}a m_{val}a. 
\label{eq:ps-fit}
\end{eqnarray}
Here $m_{val}=(m_{val,1}+m_{val,2})/2$ with $m_{val,i} (i=1,2)$ 
the bare mass of the valence quark and antiquark of the PS
meson, and $m_{sea}$ the mass of sea quark.  We mention that
details vary depending on the definitions, {\it e.g.,} terms depending only
on $m_{sea}$ are absent for the case of AWI mass since with
(\ref{eq:awi-mass}) $M_{PS}$ is zero for vanishing $m_{val}$,
and a cross term $\propto m_{val,1}a m_{val,2}a$ is
found to be necessary in the case of VWI mass.
 
The vector meson mass $M_V$ is written in a similar manner, 
adopting, however, PS meson masses as independent variables.  
We fit data with the formula  
\begin{eqnarray}
M_{V}a &=&  A + B_s \mu_{sea}a + B_v \mu_{val}a \nonumber\\
       &+&  C_s \mu_{sea}^2a^2 + C_v \mu_{val}^2a^2 + 
	C_{sv} \mu_{sea}a \mu_{val}a.    
\label{eq:vec-fit}
\end{eqnarray}
Here $\mu_{val}\!=\!(\mu_1+\mu_2)/2$ represents the average of 
PS meson mass squared
$\mu_i\!=\!M_{PS}^2(\kappa_{val,i}, \kappa_{val,i};\kappa_{sea})$  
made of a degenerate quark-antiquark pair of the valence hopping 
parameter $\kappa_{val,i}$, 
and $\mu_{sea}=M_{PS}^2(\kappa_{sea},\kappa_{sea};\kappa_{sea})$.

Fitting our hadron mass data with (\ref{eq:ps-fit}) and 
(\ref{eq:vec-fit}) we find reasonable results with 
$\chi^2/N_{\rm DF}$ in the range 0.6--2.3 
(except for (\ref{eq:ps-fit}) for VWI quark mass at $\beta=1.8$ for
which $\chi^2/N_{\rm DF}=4.0$). 
We then determine the bare lattice value of the average 
$u$ and $d$ quark mass $m_{ud}a$ 
by fixing the valence and sea quark masses to be degenerate in
(\ref{eq:ps-fit}) and (\ref{eq:vec-fit}), and requiring the experimental
value for the ratio $M_{\pi}/M_{\rho}=0.1757$. For the $s$ quark mass
$m_sa$ we use either $M_{K}/M_{\rho}=0.6477$ or $M_{\phi}/M_{\rho}=1.3267$
while keeping the sea quark mass $m_{sea}$ at the value $m_{ud}$ determined
above. The lattice scale $a^{-1}$ is set using $M_\rho=0.7684$~GeV as input.

We convert bare quark masses calculated above to renormalized quark masses
in the $\overline{MS}$ scheme at $\mu=1/a$ through $m^{\rm VWI}_Ra =
Z_m(g^2) m_q^{\rm VWI}a$ and $m^{\rm AWI}_Ra = (Z_A(g^2)/Z_P(g^2)) m_q^{\rm
AWI}a$. In these relations, the $O(a)$ improvement terms with the
coefficients $b_m$, $b_A$ and $b_P$ are also included.  For renormalization
factors and improvement coefficients, including that for $c_A$, one-loop
perturbative values for massless quark\cite{taniguchi} are used.  For the
coupling constant we adopt a mean-field improved value in the
$\overline{MS}$ scheme appropriate for the RG-improved gluon action (see
Table~\ref{tab:param} for numerical values): $g^{-2}_{\overline{MS}}(1/a) =
(3.648W_{1\times 1}-2.648W_{1\times 2})\beta/6 - 0.1006 + 0.03149N_f$ where
measured values extrapolated to zero sea quark mass are substituted for the
Wilson loops. The results for quark masses are run from $\mu=1/a$ to
$\mu=2$~GeV using the 3-loop beta function for
$N_f=2$\cite{3-loop}. Numerical values of quark masses at each $\beta$ are
given in Table~\ref{tab:quarkmass}.

\begin{figure}[htb]
\centerline{\epsfxsize=8.7cm \epsfbox{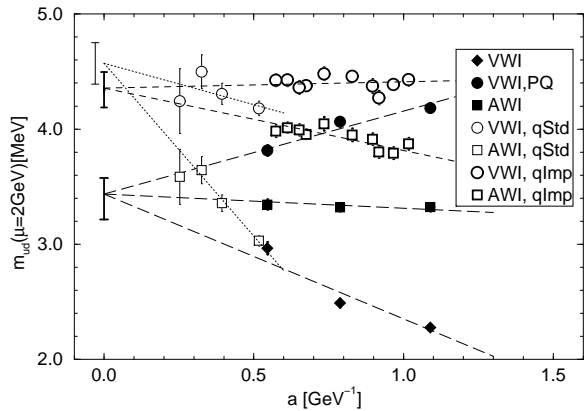}}
\vspace{-3mm}
\caption{Continuum extrapolation of the average up and down quark mass
$m_{ud}$ for full QCD (filled symbols) and quenched QCD (thick open symbols) 
obtained with the improved action, and quenched results with 
the standard action (thin open symbols). 
Lines show combined fits linear in $a$.} 
\label{fig:mud}
\end{figure}

In Fig.~\ref{fig:mud} we show our two-flavor full QCD results 
for $m_{ud}^{\overline{MS}}(2\,{\rm GeV})$ with filled symbols.  
The values for the three definitions, 
while differing sizably at
finite lattice spacings\cite{SESAMquark}, tend to converge toward the
continuum limit.  A similar trend has been seen in the 
quenched data for the standard action\cite{cp-pacs}, reproduced in 
Fig.~\ref{fig:mud} with thin open symbols.   

For our choice of the improved action, the scaling violation starts at
$O(g^2_{\overline{MS}}(\mu)a)$ for the quark masses at the scale
$\mu=2$~GeV. We therefore make a continuum extrapolation linear in $a$. The
results are given in Table~\ref{tab:quarkmass}. The masses for the three
definitions are consistent with each other at two-sigma level of
statistics.  Hence we carry out a combined linear fit, as shown in
Fig.~\ref{fig:mud} by dashed lines, obtaining $m_{ud}=3.44(9)$ with
$\chi^2/N_{\rm DF}=2.9$ where the error is only statistical. The systematic
error of the continuum extrapolation is estimated from the spread of values
obtained by separate fits of data for the three definitions.  The
fractional error thus calculated is given in Table~\ref{tab:errors}.

This table lists our estimate for two more systematic errors that we need
to incorporate. One is an uncertainty due to chiral extrapolations. We
estimate this error from the change of the combined linear fit in the
continuum limit when the quadratic term $\mu^2$ in the vector mass formula
(\ref{eq:vec-fit}) is replaced by $\mu^{3/2}$ or cubic terms $m^3$ are
included in the PS mass formula (\ref{eq:ps-fit}). Another is the error due
to the use of one-loop perturbative values for the renormalization factors.
As non-perturbative values are not yet available, we estimate the effect of
higher order contributions by recalculating masses while either shifting
the matching scale from $\mu=1/a$ to $\mu=\pi/a$ or using an alternative
definition of coupling given by $g^{-2}_{\overline{MS}}(1/a) = W_{1\times
1}\beta/6 + 0.2402 + 0.03149N_f$ using only the plaquette.

Combining the statistical error and the systematic errors listed in 
Table~\ref{tab:errors} by quadrature to obtain the total error, 
we find for our final value, 
\begin{equation}
m_{ud}^{\overline{MS}}(2 \, {\rm GeV}) = 3.44^{+0.14}_{-0.22}
\;\; {\rm MeV} \;\;\; (N_f=2).
\label{eq:ud-mass}
\end{equation}

Our full QCD result for the average $u$-$d$ quark mass is considerably lower 
than our previous quenched result for the standard action 
given in\cite{cp-pacs} as 
\begin{equation}
m_{ud}^{\overline{MS}}(2 \, {\rm GeV}) = 4.57(18) 
\; {\rm MeV} \;\; (N_f=0,\: {\rm standard}),
\end{equation}
where the error is only statistical.
In order to confirm that the decrease is a dynamical sea quark effect,
we carry out a quenched simulation using the same improved gluon and
quark actions as for the full QCD runs.  This simulation is made at 10
values of $\beta$ chosen so that the string tension matches that of 
two-flavor full QCD for each simulated value of sea
quark mass and for the chiral limit at $\beta\! =1.95$ and 2.1.  

Analyses leading from hadron masses to quark masses parallel those for full
QCD.  In particular we employ polynomial chiral expansions of the form
(\ref{eq:ps-fit})--(\ref{eq:vec-fit}), except that terms
referring to sea quark masses are dropped. As a cross-check we also make an
analysis parallel to the one in\cite{cp-pacs}, employing quenched chiral
perturbation theory formulae, and obtain consistent results. 

We plot results of this quenched analysis with thick open symbols in 
Fig.~\ref{fig:mud}.  Good consistency is observed between 
the continuum values for the standard and improved actions.  
We also note that scaling violations are visibly reduced for the latter.  
Making a combined linear extrapolation we obtain in the continuum limit,
\begin{equation}
m_{ud}^{\overline{MS}}(2 \, {\rm GeV}) = 4.36^{+0.14}_{-0.17} 
\; {\rm MeV} \; (N_f=0,\: {\rm improved}),
\end{equation}
where the error is estimated in a similar way as for full QCD. 
From this analysis we conclude that the effect of two dynamical quarks is
to decrease $m_{ud}^{\overline{MS}}(2\,{\rm GeV})$ by about 25\%.

\begin{figure}[htb]
\vspace{-5mm}
\centerline{\epsfxsize=8.7cm \epsfbox{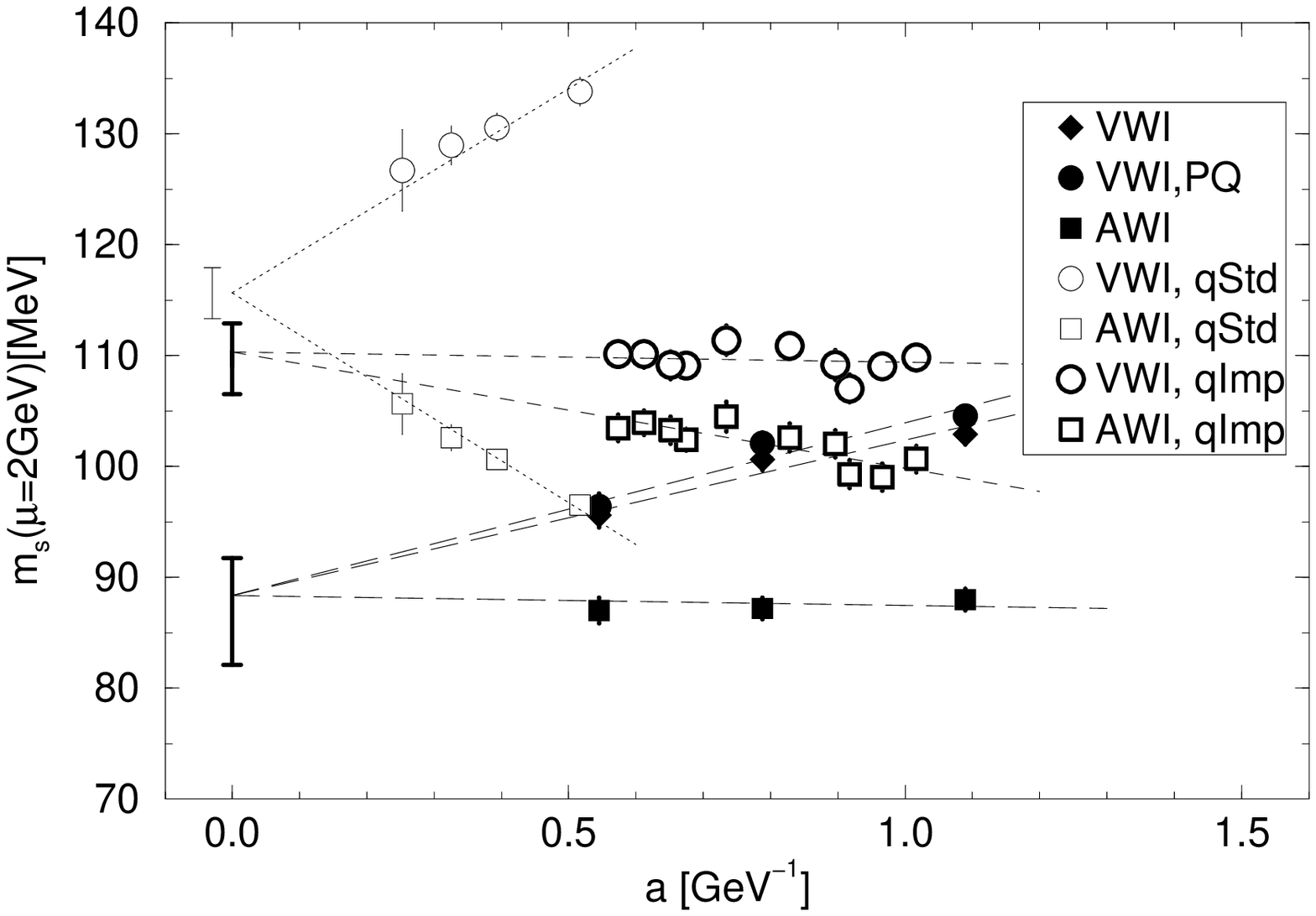}}
\vspace{-3mm}
\caption{Continuum extrapolation of the strange quark mass $m_s$ using
$M_K$ as input. Symbols have the same meaning as in Fig.~\ref{fig:mud}.}
\label{fig:ms-K}
\end{figure}

\begin{figure}[htb]
\vspace{-5mm}
\centerline{\epsfxsize=8.7cm \epsfbox{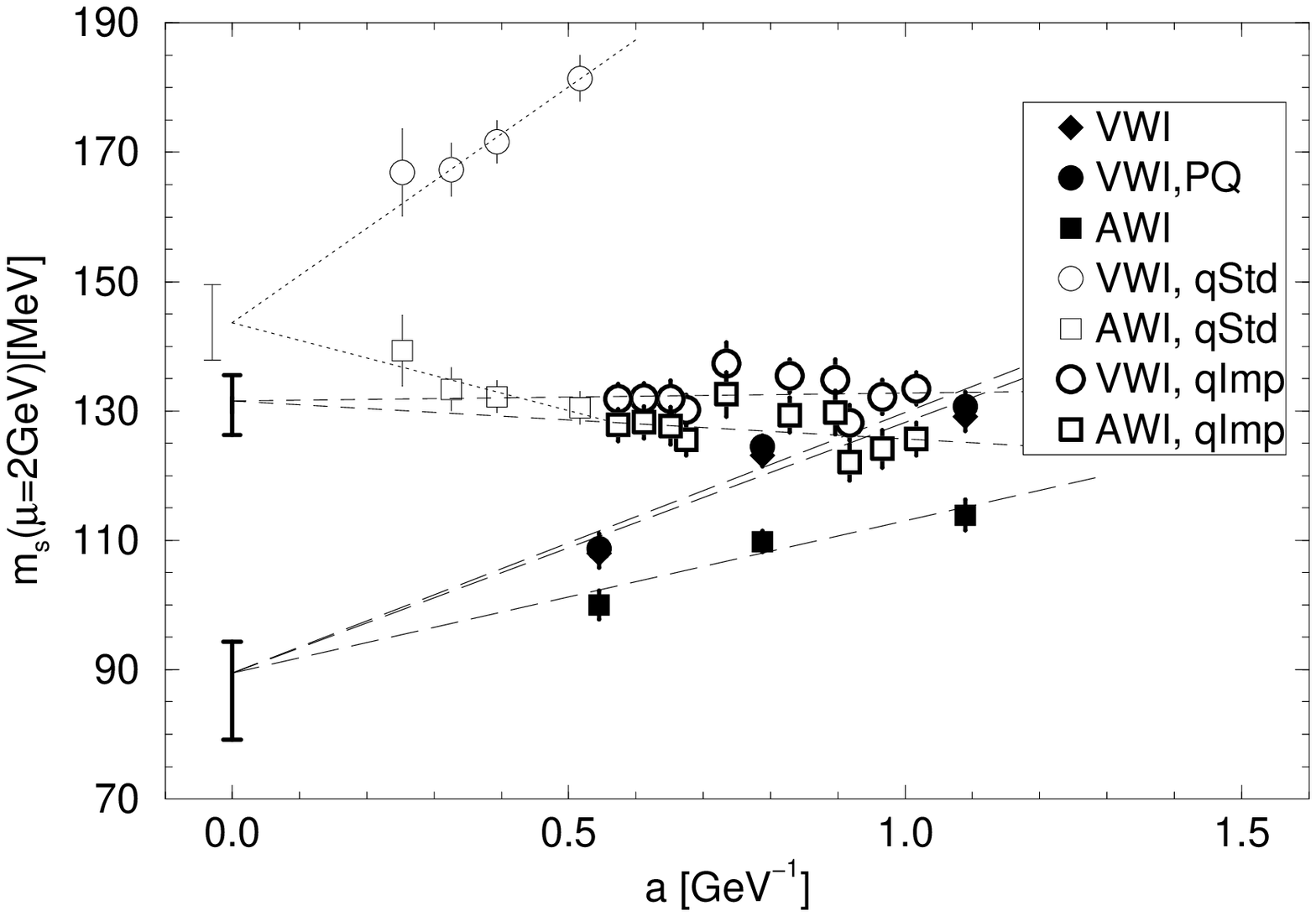}}
\vspace{-3mm}
\caption{Continuum extrapolation of the strange quark mass $m_s$ using
$M_\phi$ as input. Symbols have the same meaning as in Fig.~\ref{fig:mud}.}
\label{fig:ms-phi}
\end{figure}

In Fig.~\ref{fig:ms-K} we show results for the strange quark mass
$m_s^{\overline{MS}}(2\,{\rm GeV})$  determined from the $K$ meson mass. 
A parallel figure obtained with the $\phi$ meson mass is given in
Fig.~\ref{fig:ms-phi}. Using $K^*$ instead of $\phi$ gives the same results
within 1\%. The strange quark is heavy enough so that the
difference between $\kappa_c$ and $\kappa_c^{PQ}$ has only small effects on
the VWI masses in full QCD.  
Employing combined linear continuum extrapolations in $a$ (with
$\chi^2/N_{\rm DF}=1.3$ and 3.0, respectively), 
and estimating the error in the same way as for
$m_{ud}$ (see Table~\ref{tab:errors} for details), we obtain
\begin{eqnarray}
m_s^{\overline{MS}}(2 \, {\rm GeV}) &=& 
     88^{+4}_{-6}\;\; {\rm MeV} \;\;\; M_K  \; {\rm input} \;\; (N_f=2) 
\label{eq:smassK}\\
&=&  90^{+5}_{-11}\;\; {\rm MeV} \;\;\; M_\phi \; {\rm input} \;\; (N_f=2).
\label{eq:smassPhi}
\end{eqnarray}
With (\ref{eq:ud-mass}) this gives 
$m_s^{\overline{MS}}/m_{ud}^{\overline{MS}}=26(2)$
to be compared with 24.4(1.5)\cite{ChPT-Quarkmass} 
computed from chiral perturbation theory to one loop. 
  
Similar analyses for quenched QCD lead to the results
\begin{eqnarray}
\noalign{\hbox{$M_K$  input:}}
m_s^{\overline{MS}}(2 \, {\rm GeV}) &=& 
     110^{+3}_{-4} \: {\rm MeV} \; (N_f=0,\: {\rm improved})\\
&=&  116(3) \: {\rm MeV} \; (N_f=0,\: {\rm standard})\\
\noalign{\hbox{$M_\phi$  input:}}
m_s^{\overline{MS}}(2 \, {\rm GeV}) &=& 
     132^{+4}_{-6} \: {\rm MeV} \; (N_f=0,\: {\rm improved})\\
&=&  144(6) \: {\rm MeV} \; (N_f=0,\: {\rm standard}),
\end{eqnarray}
where the values for the standard action\cite{cp-pacs} are also quoted 
for comparison. 

The quenched values for the standard and improved
actions are mutually consistent for each choice of the input.
This confirms the existence of a systematic uncertainty of 20--30\%
in the value of $m_s^{\overline{MS}}$ in quenched QCD\cite{cp-pacs}.

One of our important results for the strange quark mass is that this
uncertainty disappears within an error of 10\% by the inclusion of two
flavors of sea quarks.  The consistency reflects a closer agreement of
the $K-K^*$ and $K-\phi$ mass splittings with experiment in our two-flavor
QCD results compared to the quenched case\cite{preliminary}.

Another important result is that dynamical quark effects reduce the value
of $m_s^{\overline{MS}}$ significantly, from the range 110--140 MeV in
quenched QCD to 90 MeV for two-flavor full QCD. It will be interesting to
see whether the inclusion of dynamical effects of the strange quark itself
would decrease the value of $m_s^{\overline{MS}}$ even further. This is an
important issue to settle as our two-flavor results are already close to
the lower bounds estimated from the positivity of spectral
functions\cite{ChPT-SumRules}. 

Clearly, establishing the values of light quark masses incorporating three
flavors of dynamical quarks will be one of the main tasks of future lattice
QCD calculations.

This work is supported in part by Grants-in-Aid of the Ministry of
Education (Nos. 09304029, 10640246, 10640248, 10740107, 11640250, 11640294,
11740162).  AAK and TM are supported by the JSPS Research for the Future
Program (No. JSPS-RFTF 97P01102). GB, SE, TK, KN and HPS are JSPS Research
Fellows.

\newpage

\begin{table}[htb]
\hsize\textwidth\columnwidth\hsize\csname
@twocolumnfalse\endcsname
\caption{\label{tab:param}
Run parameters of two-flavor full QCD simulations. The scale $a$ is set by
$\protect M_\rho=768.4$~MeV.} 
\begin{tabular}{cccccccccc}
$\beta$ & $g^{2}_{\overline{MS}}(\mu=1/a)$ & $L^3\times T$ & $c_{SW}$ 
& $a$(fm) & $La$(fm) 
& \multicolumn{4}{c}{$m_{\rm PS}/m_{\rm V}$ for sea quarks : \#traj.} \\
\tableline 
1.80 & 3.168 & $12^3{\times}24$ & $1.60$ & 0.215(2) & 2.58(3) 
& $0.807(1) \!:\! 6250$ & $0.753(1) \!:\! 5000$  & $0.694(2) \!:\! 7000$ & 
  $0.547(4) \!:\! 5250$ \\ 
1.95 & 2.816 & $16^3{\times}32$ & $1.53$ & 0.155(2) & 2.48(3) 
& $0.804(1) \!:\! 7000$ & $0.752(1) \!:\! 7000$ & $0.690(1) \!:\! 7000$ & 
  $0.582(3) \!:\! 5000$ \\ 
2.10 & 2.565 & $24^3{\times}48$ & $1.47$ & 0.108(1) & 2.58(3) 
& $0.806(1) \!:\! 4000$ & $0.755(2) \!:\! 4000$ & $0.691(3) \!:\! 4000$ & 
  $0.576(3) \!:\! 4000$ \\ 
\end{tabular}
\narrowtext
\end{table}

\begin{table}[htb]
\hsize\textwidth\columnwidth\hsize\csname
@twocolumnfalse\endcsname
\caption{\label{tab:quarkmass}
Renormalized quark masses (in MeV) in the $\overline{\rm MS}$-scheme
at $\mu=2$~GeV for each $\beta$, and in the continuum obtained by 
linear fits in $a$ (for these fits $\chi^2/N_{\rm DF}$ is also given). 
All errors are statistical.}
\begin{tabular}{cr@{.}lr@{.}lr@{.}lr@{.}lr@{.}lr@{.}lr@{.}lr@{.}lr@{.}l}
$\beta$   & 
\multicolumn{2}{c}{$m_{ud}^{\rm VWI}$}     & 
\multicolumn{2}{c}{$m_{ud}^{\rm VWI,PQ}$}      &  
\multicolumn{2}{c}{$m_{ud}^{\rm AWI}$}         & 
\multicolumn{2}{c}{$m_s^{\rm VWI}$}($K$)     & 
\multicolumn{2}{c}{$m_s^{\rm VWI,PQ}$}($K$)      &  
\multicolumn{2}{c}{$m_s^{\rm AWI}$}($K$)         & 
\multicolumn{2}{c}{$m_s^{\rm VWI}$}($\phi$)  & 
\multicolumn{2}{c}{$m_s^{\rm VWI,PQ}$}($\phi$)   &  
\multicolumn{2}{c}{$m_s^{\rm AWI}$}($\phi$)  \\
\hline
1.8  &   2&277(27) &   4&183(42) &   3&322(37)  
     & 102&92(92)  & 104&54(93)  &  88&0(1.0)  
     & 129&1(2.2)  & 130&7(2.2)  & 113&9(2.4)  \\
1.95 &   2&489(38) &   4&064(43) &   3&321(38)
     & 100&65(98)  & 102&08(99)  &  87&2(1.0)
     & 123&1(1.7)  & 124&5(1.7)  & 109&8(1.7)  \\
2.1  &   2&966(55) &   3&816(47) &   3&344(46)
     &  95&6(1.1)  &  96&4(1.1)  &  87&0(1.2)
     & 108&0(2.2)  & 108&8(2.2)  & 100&0(2.2) \\
$a\to 0$&  3&47(10)   &   3&50(10)  &  3&36(9)
     & 89&4(2.3)   &  89&5(2.3)  & 85&8(2.4)
     & 90&1(4.9)   &  90&3(4.9)  & 88&1(4.9) \\
$\chi^2/N_{\rm DF}$ & 
\multicolumn{2}{c}{10.8  }    &   
\multicolumn{2}{c}{2.4   }    &  
\multicolumn{2}{c}{0.07  }    & 
\multicolumn{2}{c}{2.1   }    &   
\multicolumn{2}{c}{2.7   }    &  
\multicolumn{2}{c}{0.03  }    &
\multicolumn{2}{c}{6.0   }    &   
\multicolumn{2}{c}{6.5   }    &  
\multicolumn{2}{c}{2.4   }
\end{tabular}
\narrowtext
\end{table}

\begin{table}[htb]
\caption{\label{tab:errors}
Contributions to total error in continuum limit.}
\begin{tabular}{ccccc}
                     & statistical & chiral   & Z-factor & cont. extrap.\\
\hline
$m_{ud}$             & $+2.6\%$ & $+1.2\%$ & $+2.3\%$ & $+1.7\%$  \\
                     & $-2.6\%$ & $-2.3\%$ & $-5.0\%$ & $-2.3\%$  \\
$m_s$ ($K$-input)    & $+2.4\%$ & $+1.6\%$ & $+2.2\%$ & $+1.4\%$  \\
                     & $-2.4\%$ & $-2.2\%$ & $-5.6\%$ & $-2.8\%$ \\
$m_s$ ($\phi$-input) & $+4.8\%$ & $+1.5\%$ & $+1.7\%$ & $+0.9\%$  \\
                     & $-4.8\%$ & $-7.6\%$ & $-6.9\%$ & $-1.6\%$  \\
\end{tabular}
\end{table}

\end{document}